\documentclass[aps,prb,twocolumn,superscriptaddress,showpacs]{revtex4}
\usepackage{graphicx}

\begin{document}

\title{Correlation effects for semiconducting single wall carbon
nanotube: a density matrix renormalization group study}

\author{Fei Ye}
\affiliation{Center for Advanced Study, Tsinghua University,
Beijing 100084, China}
\author{Bing-Shen Wang}
\affiliation{State Key Laboratory of Semiconductor Superlattice
and Microstructure
\\ and
Institute of Semiconductor, Academia Sinica, Beijing 100083,
China\\}
\author{Jizhong Lou}
\affiliation{Institute of Theoretical Physics, Academia Sinica,
Beijing 100080, China }
\author{Zhao-Bin Su}
\affiliation{Institute of Theoretical Physics, Academia Sinica,
Beijing 100080, China }
\begin{abstract}
In this paper, we report the applicability of the density matrix
renormalization group(DMRG) approach to the cylindrical single
wall carbon nanotube (SWCN) for purpose of its correlation
effect.  By applying the DMRG approach to the $t$+$U$+$V$ model,
with $t$ and $V$ being the hopping and Coulomb energies between
the nearest neighboring sites, respectively, and $U$ the onsite
Coulomb energy, we calculate the phase diagram for the SWCN with
chiral numbers ($n_{1}=3, n_{2}=2$), which reflects the
competition between the correlation energy $U$ and $V$. Within
reasonable parameter ranges, we investigate possible correlated
groundstates, the lowest excitations and the corresponding
correlation functions in which the connection with the excitonic
insulator is particularly addressed.
\end{abstract}

 \maketitle

Since the discovery of carbon nanotube, many efforts, both experimental and
theoretical, have been put into understanding the electron correlation effect,
specially for thin single wall carbon nanotubes(SWCN). The transport as well as
the angle-integrated photoemission\cite{Bockrath,Ishii}studies on the bundles of
SWCN show interesting evidence of the non-trivial correlation effects which are
believed to be contributed by the metallic SWCNs and are consistent with the
Tomonaga-Luttinger theory\cite{Balents,Krotov,Kane1}. For the semiconducting
SWCNs, the discrepancy between the experimentally measured ratio of the
absorption frequency to the emission frequency\cite{Connell,Bachilo} and that
predicted by the tight binding approximation(TBA)\cite{Saito} is also attributed
to the correlation effects in literatures\cite{Kane2,Spataru,Zhao}.
Theoretically, the correlation effect of SWCNs so far has been
treated\cite{Balents,Krotov,Kane1,Kane2,Spataru,Zhao,Ando,Chang, Sancho} by {\it
ab initio} numerical calculations, perturbative renormalization group(RG), and
various improved or modified Hartree-Fock approximations(HFA). However in these
approaches the strong correlation effects are not easy to be treated seriously.

It is known that the $\pi$-electron has its wave function extending
perpendicular to the surface of SWCN\cite{Saito} with a weak overlap among
themselves, so that its kinetic energy in a range of 2.4-3.2eV is much
smaller than the onsite Coulomb energy about 11.76eV\cite{Harrison} while
the nearest neighboring(NN) Coulomb energy is also as big as an order of
5eV. Therefore the SWCN is a system which might exhibit strong correlation
effect. We consider an extended Hubbard model to study the correlation
physics of SWCNs
\begin{eqnarray}
H&=&\sum_{<i,j>}c^\dagger_{i,\sigma}c_{j,\sigma}+h.c.+
U\sum_{i}n_{i,\uparrow}n_{i,\downarrow}
 \nonumber \\
&+&V\sum_{i,j}(n_{i}-1)(n_{j}-1)\,,
\end{eqnarray}
where the hopping energy $t$ is set to be unit, while $U$ and $V$ are onsite and
NN Coulomb energies, respectively. $c_{i}$ and $c^\dagger_{i}$ are the electron
annihilation and creation operators with $i$ the site index, respectively. The
particle number operator for spin $\sigma$ at site $i$ is
$n_{i,\sigma}=c^\dagger_{i,\sigma} c_{i,\sigma}$ and
$n_{i}=n_{i,\uparrow}+n_{i,\downarrow}$. Here we neglect the curvature induced
difference among the three hopping directions on each site. This minimal model
has been studied in Refs.[\onlinecite{Krotov,Sancho}] for metallic tubes by
perturbative RG and unrestricted HFA, respectively. It has also been applied to
the graphite system\cite{Tchougreeff}.

In this paper, we report the applicability of the DMRG approach to the
cylindrical SWCN for purpose of studying its correlation effect. By
applying this approach to the $t$+$U$+$V$ model Eq.(1), we compute the phase
diagram for the SWCN with chiral numbers ($n_{1}=3, n_{2}=2$), which essentially
reflects the competition between the correlation energies $U$ and $V$.  Within
reasonable parameter ranges of $U/t$ and $V/t$, we investigate possible
correlated groundstates, the lowest excitations and the corresponding
correlation functions.

As well accepted, the DMRG approach is an accurate and efficient treatment for
the quasi-1D strongly correlated systems\cite{SRWhite}. The main problem for the
implementation of DMRG to the SWCN lies in how to construct an appropriate
superblock configuration for such a cylindrically warped hexagonal lattice
sheet. We notice that, following the literature[\onlinecite{CTWhite}], for a
given pair of chiral numbers $(n_{1}, n_{2})$ with $N$ the greatest common
divisor, the corresponding SWCN can be constructed by successive piling of
motifs along the tube axis conducted by a screw operation\cite{CTWhite}, where
each motif contains $N$ unit cells in coincidence with the $N$-fold rotational
symmetry of the tube.  This implies an $N$-multiple helical structure hidden
in the SWCN.  The unit cells are labelled by two integers $(m,l)$ with
$l=1,2,...,N$ which corresponds the $l$-th cell in the $m$-th motif. The two
carbon atoms in one unit cell are distinguished by an index $\nu=A,B$.  Based
upon the above observation, we map such $N$-multiple helical SWCN onto a $2N$
chain lattice so as to make the DMRG procedure can be straightforwardly
implemented. In this helical coordinate system, the site $(m,l,A)$ has NNs
labelled by $(m, l, B)$, $(m-\frac{n_{1}}{N}, l+{p_{1}}, B)$,
$(m+{\frac{n_{2}}{N}}, l-{p_{2}}, B)$, where $p_{1},p_{2}$ are integer solutions
of equation $n_{1}p_{2}-n_{2}p_{1}=N$\cite{CTWhite}. In particular, for case of
$N=1$, which is the most advantaged, all the unit cells on the tube can be
threaded by only one helical curve and $l$ becomes a dummy index.  Then the
Hamiltonian in Eq.(1) can be rigorously mapped to a twisted two-chain model with
finite long range hoppings as $H=H_{t}+H_{U}+H_{V}$ with
\begin{eqnarray}
    H_t&=&\sum_{m,\mu\in\mathcal{S}}\sum_{\sigma}{c}^{A^{\dag}}_{m,\sigma}{c}^{B}_{m+\mu,\sigma}+h.c.  \nonumber \, \\
    H_U&=&U\sum_{m,\nu=A,B}n^{\nu}_{m,\uparrow}n^{\nu}_{m,\downarrow}  \nonumber \, \\
    H_V&=&V\sum_{m,\mu\in\mathcal{S}}(n^{A}_{m}-1)(n^{B}_{m+\mu}-1) \, \label{twochain}
\end{eqnarray}
where $\mathcal{S}=\{0,-n_1,n_2\}$,
$n^\nu_{m,\sigma}={c}^{\nu^{\dag}}_{m,\sigma}{c}^{\nu}_{m,\sigma}$,
$n^\nu_{m}=n^{\nu}_{m,\uparrow}+n^{\nu}_{m,\downarrow}$ and
${c}^{\nu}_{m,\sigma}$ is exactly the same operator $c_{i,\sigma}$ but with its
index relabelled in terms of motif-cell labels.  Fig.~1a and 1b show such a
mapping for the tube $(n_1=3,n_2=2)$ with $N=1$. 
\begin{figure}[tbh]
\centering
\begin{minipage}[c]{0.5\textwidth}
\scalebox{0.8}[0.8]{\includegraphics{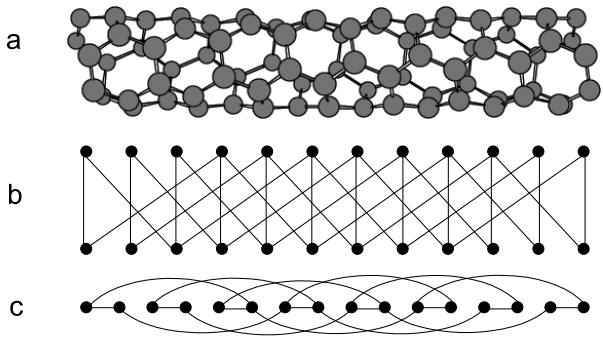}}
\end{minipage}
\caption{Illustration for the mapping from a cylindrical SWCN(1a)
with index (3,2) to a twisted two chain model with finite long
range hopping (1b), where the solid lines connect the NN pairs.
For purpose of DMRG calculation, we further squeeze the two chains
in (1b) into a single chain (1c) with $A$ atom and $B$ atom
arranged alternatively, where the lines  are kept with the same
meaning as in (1b).}
\end{figure}

To extract most physical implications with small enough computing effort, in
this paper, we apply the standard DMRG approach to the intrinsic semiconducting
SWCN $(3, 2)$ with $N=1$, in which the open boundary condition(OBC) is engaged.
The computed groundstate phase diagram is plotted in Fig.~2. It can be divided
into three phases as the \emph{excitonic insulator}($E$), \emph{normal
semiconductor}($N$), and \emph{Mott insulator}($M$), as will be clear shortly.

For our minimal model as Eq.(2) when both $U$ and $V$ are small, the system is a
semiconductor which can be divided into two regions $N_s$ and $N_T$ according to
the first excitations.  
\begin{figure}[tbh]
\centering
\begin{minipage}[c]{0.5\textwidth}
\scalebox{0.65}[0.65]{\includegraphics{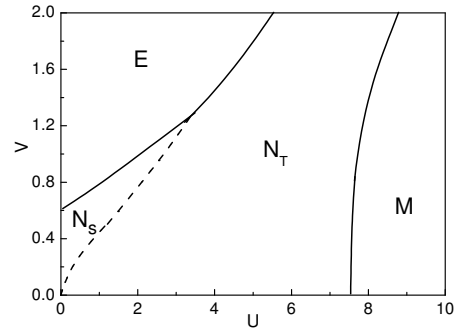}}
\end{minipage}
\caption{The phase diagram of semiconducting SWCN(3,2).  See the text
for details.}
\end{figure}
The boundary between them is shown as the dashed line in Fig.~2, across which a
level crossing takes place for the gapped excited states with different spin
symmetries, i.e., the singlet in the $N_{s}$ region and the triplet in the
$N_{T}$ region, which are favored by $V$ and $U$, respectively.  We interpret
these excitations in connection with the singlet and triplet excitons for the
intrinsic semiconducting SWCN.  The groundstate in these two regions is a
non-degenerate singlet state, of which the equal time correlation function(ETCF)
is exponentially decaying in both spin and charge sectors.  The difference is
that density-density correlation is much stronger than the spin-spin correlation
in the $N_s$ region and vice versa in the $N_T$ region.

If we still keep $U$ small but increase $V$, the first spin singlet excitation
would get lowered. When it reaches the groundstate, the system then undergoes a
transition into a new phase in association with the excitonic
insulator\cite{Jerome}. As a consequence of the ``fallen state", there are two
mutually orthogonal degenerate groundstates in the $E$ phase as
$|G^{(E)}_+\rangle$ and $|G^{(E)}_-\rangle$, in which the 2-fold degeneracy is
understood in the thermodynamic limit. Both of them are spin singlet and
eigenstates of the parity, belonging to eigenvalues $\pm1$, respectively.
Moreover, the amplitude of the density-density correlation increases with
increasing $V$ as shown in the upper panel of Fig.~3, until into the $E$ phase
it is identical for both the two degenerate groundstates and no longer decays.
This non-decaying correlation could be interpreted as the existence of a charge
density wave in the two groundstates. When computing the matrix element of the
particle number operator $n_{m}^{\nu}$ between these two degenerate states, it
is found that the diagonal term $\langle G^{(E)}_+|n_{m}^{\nu}|G^{(E)}_+\rangle$
(or$\langle G^{(E)}_-|n_{m}^{\nu}|G^{(E)}_-\rangle$) is trivially equal to 1,
but the off-diagonal matrix element is
\begin{eqnarray}
\langle G^{(E)}_+|n_{m}^{\nu}|G^{(E)}_-\rangle\sim
(-1)^{\nu}\tilde{n}
\end{eqnarray}
with $\tilde{n}$ being a site-independent constant and $(-1)^{\nu}=\pm 1$ for
$\nu=A,B$, respectively. By mixing the two states $|G^{(E)}_\pm\rangle$, one can
construct the groundstate with alternative density distribution in the $A$ and
$B$ sublattices. This can be viewed as the exciton(particle-hole pair)
condensation in the groundstate of $E$ phase with a spontaneous symmetry
breaking between the sublattices.  
\begin{figure}[tbh]
\centering
\begin{minipage}[c]{0.5\textwidth}
\scalebox{0.75}[0.75]{\includegraphics{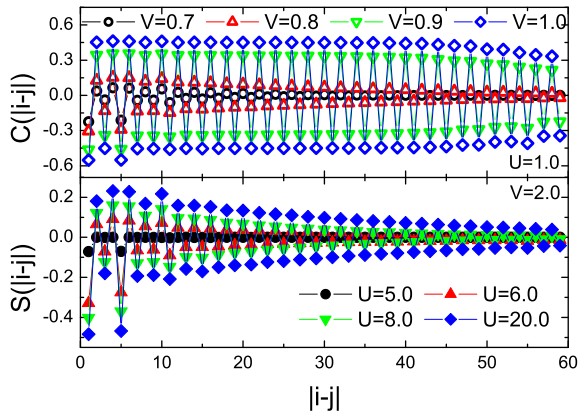}}
\end{minipage}
\caption{The charge and spin correlation functions. The coordinate system for
the horizontal axis is explained in Fig.~1c. The origin is set on the atom $A$
in the middle cell of the tube. The data are obtained by keeping 2000 states
and with 1-2 finite size sweep to get convergent results. The upper panel gives
the change of charge correlation $C(|i-j|)$ with respect to $V$ for fixed
$U=1.0$ and the lower panel is the antiferromagnetic spin correlation $S(|i-j|)$
for different $U$ with fixed $V=2.0$.}
\end{figure}

Starting from the region $N_T$, if we further increase $U$, the triplet first
excitation would be lowered continuously with the groundstate spin-spin ETCF
growing. In the lower panel of Fig.~3, we show the variation of spin
correlation versus $U$ for fixed value of $V$.  Until some critical value of
$U=U_{c}$, the triplet spin gap vanishes in the thermodynamic limit\cite{mott},
and a transition from phase $N$ to the Mott phase $M$ takes place. In the Mott
phase, besides a finite charge gap the system develops gapless spin wave type
excitations. The Mott phase is far from the possible realistic parameter region
of $U$ and $V$ for the SWCN(3,2), however it may be realized in the small gap
semiconductor or metallic tube.

The phase boundaries in Fig.~2 are scanned by computing the energy gap in the
particle-hole channel for various $U$ and $V$.  In this procedure, we adopt the
infinite algorithm with 1000 states being kept and then extrapolate the finite
size data to the thermodynamic limit.  The calculated phase boundaries will also
be checked by other independent method below. In Fig.~4, we give the variation
of the gap with respect to $U$ for some fixed values of $V$ as examples.

For $V=2$ in Fig.~4a, the 2-fold degeneracy of the groundstates in $E$ phase is
lifted abruptly at the $E-N_T$ phase boundary $U_{C}=5.6$ and a finite spin
triplet excitation gap is opened. It decreases with increasing $U$ and vanishes
at about $U=8.8$.  Fig.~4b with $V=0.8$ shows a different picture. The singlet
gap emerges gradually from zero at the $E-N_s$ phase
\begin{figure}[tbh]
\centering
\begin{minipage}[c]{0.5\textwidth}
\scalebox{0.7}[0.7]{\includegraphics{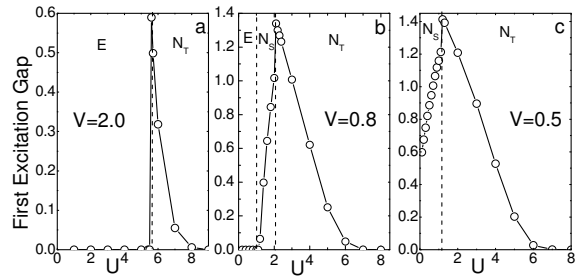}}
\end{minipage}
\caption{The first excitation gap as a function of onsite Coulomb potential $U$,
for $V=2.0$ in (4a), $V=0.8$ in (4b),  and $V=0.5$ in (4c).  The vertical dashed
lines indicate the transition points.}
\end{figure}
boundary.  When the increasing $U$ reaches the $N_{s}-N_T$ boundary, a level
crossing takes place from the singlet excitation to  the triplet excitation.
Fig.~4c with $V = 0.5$ only shows a $N_{s}-N_T$ transition qualitatively the
same as that in Fig.~4b.

To further explore the nature of the above shown phase boundaries, and in
particular, as an independent check, we compute the average of double occupancy
and the NN density correlation over the groundstate, denoted by $\delta$ and
$\rho$, respectively, as in Ref.[\onlinecite{Jeckelmann}]
\begin{eqnarray}
\delta&\equiv&\partial E_{GS}/\partial U=(H_U/UL)\nonumber\\
\rho&\equiv&\partial E_{GS}/\partial V=(H_V/VL)\;.
\end{eqnarray}
Here $E_{GS}$ is the groundstate energy per site and the Hellmann-Feynmann
theorem is applied.  For fixed $V=2.0$ or $0.8$, the $\delta$ and $\rho$ as
functions of $U$ are plotted in Fig.~5. The calculations are performed on the
tube with 120, 160  and 200 carbon atoms by keeping 2000 states and sweeping 1-2
times to get convergent results. The results calculated from different atom
numbers coincide with one another nicely, which provides a strong evidence
that the calculated main features of the  transition will survive in the
thermodynamic limit. 

As shown in the right panel in Fig.~5, there is an abrupt jump for both $\delta$
and $\rho$ at the critical point $V=2.0$ and $U=5.6$. It indicates the phase
boundary between $E$ and $N_T$ belongs to the first order transition.  Note that
the phase boundaries(dashed lines) in Fig.~5 are extracted from the phase
diagram Fig.~2, which agree with the data of $\delta$ and $\rho$ very well.  If
we further increase $U$ from region $N_T$ to phase $M$, $\delta$ and $\rho$ vary
smoothly across the phase boundary(see also right panel of Fig.~5). It is
understood that the $N_T-M$ phase transition is of the Kosterlitz-Thouless type
without symmetry breaking, where the $N_T$ region with a gapful (massive)
triplet excitation spectrum transits into the $M$ phase with gapless(massless)
spin excitations. In the left panel of Fig.~5, there is a discontinuity in the
slope of both $\delta$ and $\rho$ as functions of $U$ on the $N_s-E$ phase
boundary, which means this boundary belongs to a second order phase transition.
Across the $N_s-N_T$ division line, a level crossing of the excited states, not
a phase transition, takes place.
\begin{figure}[tbh]
\centering
\begin{minipage}[c]{0.5\textwidth}
\scalebox{1.0}[1.0]{\includegraphics{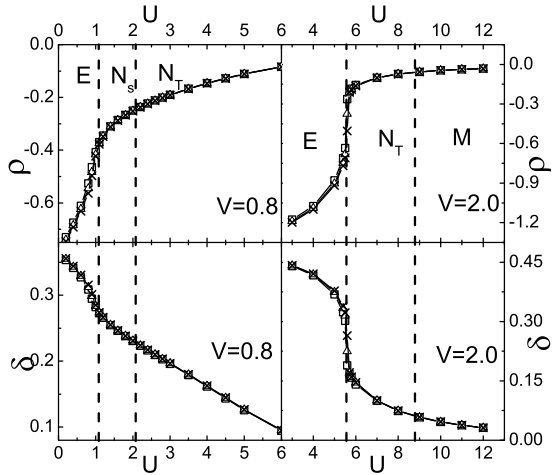}}
\end{minipage}
\caption{ The plot of $\delta$ and $\rho$ as functions of $U$ for fixed $V$. The
vertical dashed lines indicate the transition points which are extracted from
the phase diagram Fig.~2. The calculated data are for 120 atoms(open squares),
160 atoms(open triangles), and 200 atoms(cross symbol),  respectively. }
\end{figure}

Briefly the transitions from the semiconductor into the excitonic insulator and
Mott insulator take place whenever the band gap is overwhelmed completely by by
$V$ and $U$, respectively. For other semiconducting SWCNs, the augment of the
tube diameter will reduce the TBA band gap, which will consequently shift the
phase boundaries $E-N_s$ and $N_T-M$. However the phase boundaries $E-N_T$ and
$N_s-N_T$ should not change too much because they can be viewed as the balance
position of the competition between $U$ and $V$.
   
In our minimal model, the longer range(LR) part of Coulomb interaction is
ignored, which is nonessential in describing the correlated groundstate of
semiconducting SWCNs. Since we are dealing with the low energy excitations of
the intrinsic semiconducting SWCN, the main effects of the LR Coulomb
interaction are confined in the forming particle-hole excitations and its
remnant effect leads to a renormalization of $U$ and $V$.  Based on the above
argument, the groundstate of SWCN may be an excitonic insulator, normal
semiconductor with different type of first excitations and Mott insulator, which
is dependent on the chiral structure of SWCN and also the environment.  In
experiments, people can realize the transition of the correlated groundstate of
SWCNs by applying a uniform pressure or by changing the band gap via the
magnetic flux\cite{ajiki,lu,science}.

Acknowledgement: We thank sincerely to Profs. S.J.Qin, E.Tosatti,
T.Xiang and L.Yu for beneficial discussions.

\end{document}